\documentclass[preprint,preprintnumbers,amsmath,amssymb]{revtex4}

\usepackage{graphicx}% Include figure files
\usepackage{dcolumn}% Align table columns on decimal point
\usepackage{bm}% bold math

\newcommand{\be}{\begin{equation}}
\newcommand{\ee}{\end{equation}}
\newcommand{\bea}{\begin{eqnarray}}
\newcommand{\eea}{\end{eqnarray}}

\newcommand{\rme}{{\rm{e}}}
\newcommand{\rmi}{{\rm{i}}}
\newcommand{\sigmax}{{e_1}}
\newcommand{\sigmay}{{e_2}}
\newcommand{\sigmaz}{{e_3}}
\newcommand{\ssI}{P}

\begin{document}

\title{An improved formalism for the Grover search algorithm}  
\author{James M. Chappell$^{\text{*\textit{a},\textit{b}}}$, M.~A.~Lohe$^{\textit{b}}$, Lorenz von Smekal$^{c}$, 
Azhar Iqbal$^{\text{\textit{a}}}$ and Derek Abbott$^{\text{\textit{a}}}$}
\affiliation{$^{\text{\textit{a}}}$School of Electrical and Electronic Engineering,
University of Adelaide 5005, Australia \\
$^{\text{\textit{b}}}$School of Chemistry and Physics, University of
Adelaide, South Australia 5005, Australia\\
$^{\text{\textit{c}}}$  Institut f\"ur Kernphysik, Technische Universit\"at Darmstadt, Schlossgartenstra{\ss}e 9, 64289 Darmstadt, Germany \\
$^{\ast }$\texttt{Email}:~\texttt{james.m.chappell@adelaide.edu.au}}

\begin{abstract}
The Grover search algorithm is one of the two key algorithms in the field of quantum computing, and hence it is of significant interest to describe it in the most efficient mathematical formalism. We show firstly, that Clifford's formalism of geometric algebra, provides a significantly more efficient representation than the conventional Bra-ket notation, and secondly, that the basis defined by the states of maximum and minimum weight in the Grover search space, allows a simple visualization of the Grover search as the precession of a spin-$\frac{1}{2} $ particle.  Using this formalism we efficiently solve the exact search problem, as well as easily representing more general search situations.  
\end{abstract}

%\pacs{03.67.Lx}% PACS, the Physics and Astronomy Classification Scheme.

\keywords{Grover search, quantum algorithms, geometric algebra}

\maketitle

\section{Introduction}

The Grover search algorithm \cite{Grover:1996,Grover:1998,grover2001schrödinger}, seeks to evolve a wave function, from some starting state $ | \sigma \rangle $, into the solution state $ | m \rangle $, representing the set of all solutions, which upon measurement, will yield one element from this set \cite{GroverOriginal,NielsenChuang:2002,ng2002introduction}. In order to analyze this evolution, typically an orthonormal basis $ |m \rangle $ and $ |m^{\perp} \rangle $ is defined, as shown on Fig (\ref{visualGroverIteration}) upon which the starting state $ | \sigma \rangle $ is plotted.
However in this paper we use an alternative basis, the states of maximum and minimum weight, which allows the initial sate $ | \sigma \rangle $ and the solution state $ | m \rangle $ to be symmetrically positioned in this space, allowing the conceptualizing of the Grover search process, analogous to the precession of a spin-$\frac{1}{2}$ particle in a magnetic field, precessing in this case, from the direction of the initial state $ | \sigma \rangle $  to the solution state $ | m \rangle $ \cite{CIL}. This approach is similar to an SO(3) picture which has previously been developed  \cite{GuiLong}, which also plots the path of the state vector during the application of the Grover operator.  Clifford algebra has also been applied previously to Grover's algorithm \cite{alves2010clifford,somaroo1998expressing,gregoric2009quantum}, however the approach adopted here combines the benefits of an efficient representation as well as an integral geometric visualization.  

\subsection{The standard Grover search}

Given a search space of $ N $ elements, with $ M $ of these elements being solutions to a search query as defined by an oracle $ f(x) $, where by definition, $ f(x) = 1 $ if $ x $ is a solution, and $ f(x) = 0 $ if $ x $ is not a solution, we
set up two equivalence classes defined by the following two states:
\bea
| m \rangle & = &  \frac{1}{\sqrt{M}} \sum_{x \in M} | x \rangle, \\ \nonumber
| m^{\perp} \rangle & = & \frac{1}{\sqrt{N-M}}  \sum_{x \notin M} | x \rangle,  \nonumber
\eea
allowing us to define a uniform superposition starting state in terms of these two states as 
\be
| \sigma \rangle = \frac{1}{\sqrt{N}} \sum_{x=0}^{N-1} | x \rangle = \sqrt{\frac{N-M}{N}} | m^{\perp} \rangle + \sqrt{\frac{M}{N}} | m \rangle .
\ee

\setlength{\unitlength}{1mm}
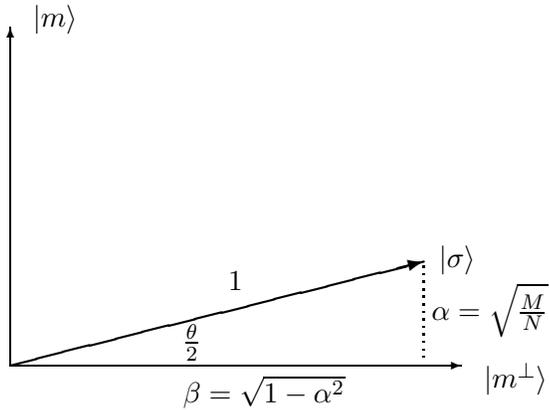
\begin{figure}[htb]
\begin{picture}(85,60)
\put(10,7){\vector(1,0){60}} \put(10,7){\vector(0,1){45}}
\thicklines \put(10,7){\vector(4,1){55}}
\put(13,52){$ | m \rangle $} \put(73,4){$ | m^{\perp} \rangle $}
\put(66,13){$ \alpha = \sqrt{\frac{M}{N}} $}
\put(67,20){$ | \sigma \rangle $} 
\put(33,9){$ \frac{\theta}{2} $} \put(39,17){1} 
\put(33,2){$ \beta = \sqrt{1-\alpha^2} $}
\multiput(65,8)(0,1){13} {\line(0,1){0.2}} 
\end{picture}
\caption{Geometry of starting state $ | \sigma \rangle $.\label{visualGroverIteration}}
\end{figure}

Grover's solution to the search process \cite{GroverOriginal}, involves iteratively applying a unitary operator $ G $ defined by
\be \label{eq:StdGrover} 
G = -G_{\sigma} G_m  = -(I - 2 | \sigma \rangle \langle \sigma | )(I - 2 | m \rangle \langle m | ).
\ee
This operator applied to the $ n = \lceil \log N \rceil $ qubits representing the search space, rotates the state vector an angle given by
\be \label{eq:sinTheta}
\sin \frac{\theta}{2} = \sqrt{\frac{M}{N}}
\ee
at each application, and after
\be \label{eq:GroverTime} 
R \le \left \lceil \frac{\pi}{4}  \sqrt{\frac{N}{M} } \right \rceil
\ee
iterations, the wave function will lie close to the solution state $ | m \rangle $.  

\subsection{Modified basis vectors for the search space}

The Grover search space, is found to be isomorphic to an su(2) space \cite{genGrover}, and using the states of maximum and minimum weight, we can find a geometric picture of the search process, in a real three-dimensional space.
Working from the two-dimensional complex space shown in Fig (\ref{visualGroverIteration}), using the well defined states $ |m \rangle $ and $ |\sigma \rangle $, we have four possible operators: $ | m \rangle \langle \sigma | $, $ | \sigma \rangle \langle m | $, $ | m \rangle \langle m | $ and $ | \sigma \rangle \langle \sigma | $.  From these we define:
\bea \label{eq:Js} 
K & = & - \frac{\beta^2}{2}\ssI \\
J_1 & = & \frac{\ssI - |\sigma \rangle \langle \sigma | - | m \rangle \langle m |}{2 |\alpha|}\\ 
J_2 & = & \frac{-\rmi \left (\alpha^* | m \rangle \langle \sigma | - \alpha | \sigma \rangle \langle m | \right )}{2 | \alpha| \beta }  \\ 
J_3 & = & \frac{|\sigma \rangle \langle \sigma | - | m \rangle \langle m |}{2 \beta},  \nonumber
\eea
where $ \alpha = \langle \sigma | m \rangle $, which may be complex, $ \beta = \sqrt{1-|\alpha|^2} $, and $ {\rm{i}} = \sqrt{-1} $ with
\be \label{eq:ProjectorId} 
\ssI = \frac{ (|\sigma \rangle \langle \sigma | - | m \rangle \langle m |)^2}{\beta^2}.
\ee
We then find 
\bea
{\rm{[}} J_i , J_j {\rm{]}} & = & \rmi \epsilon_{ijk} J_k \\ \nonumber
{\rm{[}} J_i , J_j {\rm{]}}_+ & = & \delta_{ij} \frac{P}{2} \\ \nonumber
{\rm{[}} K , J_i {\rm{]}} & = & 0, \nonumber
\eea
where $ \delta $ is the Kronecker delta symbol, and $ \epsilon $ is the Levi-Civita symbol, confirming we have an su(2) algebra.
Squaring the generators we find
\be
J_1^2 = J_2^2 = J_3^2 = \frac{1}{4 \beta^2} (|\sigma \rangle \langle \sigma | -| m \rangle \langle m |)^2 = \frac{1}{4} \ssI.
\ee
We can easily check, $ \ssI | \sigma \rangle = | \sigma \rangle $ and $ \ssI | m \rangle = | m \rangle $, with $ \ssI^2 = \ssI $ and the Casimir invariant
\bea
C = J_1^2 + J_2^2 + J_3^2 = \frac{3}{4} \ssI, \nonumber
\eea
which corresponds to a spin $ \frac{1}{2} $ system.
We have the raising and lowering operators
\be
J_{\pm}=J_1 \pm \rmi J_2 \nonumber
\ee
and requiring $ J_+ |\uparrow \rangle = 0 $ and $ J_- |\downarrow \rangle  = 0 $, we find the states of highest and lowest weight
\bea \label{maxminWeight}
|\uparrow \rangle & = & \sec \frac{\theta}{2} \left ( \sin \frac{\theta}{4} | m \rangle - \rme^{\rmi \delta} \cos \frac{\theta}{4} | \sigma \rangle \right ) \\ \nonumber
|\downarrow \rangle & = & \sec \frac{\theta}{2} \left ( \cos \frac{\theta}{4} | m \rangle -\rme^{\rmi \delta} \sin \frac{\theta}{4} | \sigma \rangle \right ), \nonumber
\eea
where $ \sin \frac{\theta}{2} = |\alpha| $ and $ \alpha = |\alpha| \rme^{\rmi \delta} $.
We then find $ J_3 |\uparrow \rangle  =  +\frac{1}{2} |\uparrow \rangle $ and $ J_3 |\downarrow \rangle  =  -\frac{1}{2} |\downarrow \rangle $, as expected for a spin$-\frac{1}{2} $ system.
Writing $ | \sigma \rangle $ and $ | m \rangle $ in this new basis we obtain
\bea \label{SpinorsInNewBasis}
|\sigma \rangle & = & \rme^{-\rmi \delta} \left (-\cos \frac{\theta}{4} |\uparrow \rangle + \sin \frac{\theta}{4} |\downarrow \rangle \right)  \\ \nonumber
|m \rangle & = & -\sin \frac{\theta}{4} |\uparrow \rangle + \cos \frac{\theta}{4} |\downarrow \rangle \\ \nonumber
|m^{\perp} \rangle & = & \cos \frac{\theta}{4} |\uparrow \rangle + \sin \frac{\theta}{4} |\downarrow \rangle. \nonumber
\eea
Using these results, we can substitute back into the Grover iteration Eq.~(\ref{eq:StdGrover}) to find
\be
G  = -I + 2 \cos^2 \frac{\theta}{2}| \uparrow \rangle \langle \uparrow | + \sin \theta | \uparrow \rangle \langle \downarrow | - \sin \theta | \downarrow \rangle \langle \uparrow |+ 2 \cos^2 \frac{\theta}{2} | \downarrow \rangle \langle \downarrow | , 
\ee
which can be immediately written in matrix form as
\be
G = \begin{bmatrix}  \cos \theta & \sin \theta \\ -\sin \theta & \cos \theta \end{bmatrix},
\ee
which shows as expected, that the Grover operation rotates the state vector by an angle $ \theta $, where the starting state will be for this basis
\be
|\sigma \rangle = \rme^{-\rmi \delta} \begin{bmatrix}  -\cos \frac{\theta}{4} \\ \sin \frac{\theta}{4}  \end{bmatrix}.
\ee

\subsection{Clifford's algebra of three-space}

Using the orthonormal basis $ | \uparrow \rangle$, $ | \downarrow \rangle $, defined in Eq.~(\ref{maxminWeight}), we now model the search process using the real associative algebra of GA.
We define unit algebraic elements $ \sigmax, \sigmay, \sigmaz $, such that $ e_1^2 = e_2^2 =e_3^2 = 1 $, and for distinct $ i $ and $ j $ we have the anticommutation rule $ e_i e_j = - e_j e_i $ \cite{Doran:2003}. The algebraic elements $ e_1,e_2,e_3 $, define a three-dimensional space, and so we can define two vectors $ \mathbf{a} = a_1 e_1 + a_2 e_2 + a_3 e_3 $ and $ \mathbf{b} = b_1 e_1 + b_2 e_2 + b_3 e_3 $, then using the distributive law of multiplication over addition we find
\bea \label{GeometricProduct}
\mathbf{a} \mathbf{b} & = & (a_1 e_1 + a_2 e_2 + a_3 e_3)(b_1 e_1 + b_2 e_2 + b_3 e_3) \\ \nonumber
& = & a_1 b_1 + a_2 b_2 + a_3 b_3 + (a_2 b_3 - b_2 a_3 ) e_2 e_3 + (a_1 b_3 - a_3 b_1 ) e_1 e_3 + (a_1 b_2 - b_1 a_2 ) e_1 e_2 \\ \nonumber
 & = & \mathbf{a} \cdot \mathbf{b} + \iota \mathbf{a} \times \mathbf{b} = \mathbf{a} \cdot \mathbf{b} + \mathbf{a} \wedge \mathbf{b}   ,  \nonumber
\eea
where $ \mathbf{a} \cdot \mathbf{b} $ is therefore the conventional dot or inner product and  $ \mathbf{a} \wedge \mathbf{b} $ is the wedge or outer product. In three dimensions we have the relationship with the conventional vector product that $ \mathbf{a} \wedge \mathbf{b} = -\iota \mathbf{a} \times \mathbf{b} $, where we have defined the trivector $ \iota = \sigmax \sigmay \sigmaz $, which represents a signed unit volume. 

Using the product defined Eq.~(\ref{GeometricProduct}), with orthonormal basis elements, we find
\be
e_i e_j = e_i.e_j + e_i \wedge e_j = \delta_{ij}+\iota \epsilon_{ijk} e_k,
\ee
indicating that we have an isomorphism between the basis vectors $ e_1, e_2, e_3 $ and the Pauli matrices through the use of the geometric product. We find that $ \iota^2 = \sigmax \sigmay \sigmaz \sigmax \sigmay \sigmaz = -1 $ and we also find that $ \iota $ commutes with all other elements of the algebra and so can be used in place of the unit imaginary $ \rmi = \sqrt{-1} $. The bivectors also square to negative one, that is $
(e_i e_j)^2 = (e_i e_j)(e_i e_j)=-e_i e_j e_j e_i=-1 $, which are used to define rotations in the plane of the bivector.

\subsubsection{Rotations in 3-space with geometric algebra}

The Grover search process involves the incremental rotation of the state vector and, in geometric algebra, in order to rotate a vector by an angle $ | \mathbf{a}| $ about an axis given by the vector $ \mathbf{a} $ we define a rotor
\be
R = \rme^{-\iota \mathbf{a}/2} = \cos (|\mathbf{a}|/2) - \iota \frac{\mathbf{a}}{|\mathbf{a}|} \sin(|\mathbf{a}|/2),
\ee
which acts by conjugation to rotate vector $ \mathbf{v} = v_1 e_1 + v_2 e_2 + v_2 e_3 $, using
\be
\mathbf{v}^{'}= R \mathbf{v} R^{\dagger} = \rme^{-\iota \mathbf{a}/2} \mathbf{v} \rme^{\iota \mathbf{a}/2} .
\ee
The $ \dagger $  is also called the {\it{reversion}} operation, which flips the order of the terms and switches the sign of $ \iota $.
The bilinear transformation used to calculate rotations is preferred, because it applies completely generally, to not only rotating vectors, but also any components of the algebra, and also in any number of dimensions.

\subsubsection{Representing quantum states in Geometric Algebra(GA)}

We can identify a simple 1:1 mapping from complex spinors to the bivectors of GA as follows \cite{Doran:2003,Venzo2007,DoranParker:2001} 
\be \label{eq:spinorMapping} 
|\psi\rangle = z_1 | \uparrow \rangle + z_2 | \downarrow \rangle = \begin{bmatrix}  a_0+\rmi a_3 \\ -a_2+ \rmi a_1 \end{bmatrix} \leftrightarrow \psi = a_0 + a_k \iota e_k.
\ee
This maps spinors to the even sub subalgebra of GA, which is closed under multiplication.

Converting the complex spinors defined in Eq.~(\ref{SpinorsInNewBasis}),  we find using Eq.~(\ref{eq:spinorMapping})
\bea
| \sigma \rangle & \mapsto & -\cos \frac{\theta}{4} -\sin \frac{\theta}{4} \iota \sigmay = - \rme^{\iota \sigmay \theta/4} \\ \nonumber
| m \rangle & \mapsto & -\sin \frac{\theta}{4} -\cos \frac{\theta}{4} \iota \sigmay = - \rme^{\iota \sigmay (\pi/2-\theta/4)} = - \iota \sigmay \rme^{-\iota \sigmay \theta/4}\\ \nonumber
| m \rangle^{\perp} & \mapsto &  \cos \frac{\theta}{4} -\sin \frac{\theta}{4} \iota \sigmay = \rme^{-\iota \sigmay \theta/4}. \nonumber
\eea
We can now transform GA type spinors into a real space representation through the transformation
\be \label{eq:polarisationVectorGA}
 S  = \psi  \sigmaz  \psi^{\dagger} ,
\ee
which gives us the three-space vectors
\bea \label{eq:mainVectors}
\sigma & = & \rme^{\iota \sigmay \theta/4} \sigmaz \rme^{-\iota \sigmay \theta/4} =  \rme^{\iota \sigmay \theta/2} \sigmax = -\sin \frac{\theta}{2} \sigmax +\cos \frac{\theta}{2} \sigmaz  \\ \nonumber
m & = &  \rme^{\iota \sigmay (\pi/2-\theta/4)} \sigmaz \rme^{-\iota \sigmay (\pi/2-\theta/4)} = -\sigmaz \rme^{\iota \sigmay \theta/2} = \rme^{\iota \sigmay (\pi - \theta/2)} \sigmaz = -\sin \frac{\theta}{2} \sigmax -\cos \frac{\theta}{2} \sigmaz  \\ \nonumber
m^{\perp} & = &  \rme^{-\iota \sigmay \theta/4} \sigmaz \rme^{\iota \sigmay \theta/4}  = \sigmaz \rme^{\iota \sigmay \theta/2} = \sin \frac{\theta}{2} \sigmax +\cos \frac{\theta}{2} \sigmaz.  \nonumber
\eea
Hence the vectors $ \sigma  $, $  m  $ and $  m^{\perp} $, can now be plotted in a real Cartesian space analogous to the Bloch sphere representation as shown in Fig (\ref{GA3Space}), and can be compared with \cite{GuiLong}. As can be seen we use the $\sigmaz $ ($z$-axis) from which to measure the angle $ \theta $,  and $ \phi $ is measured from $ \sigmax $.  

\setlength{\unitlength}{1mm}
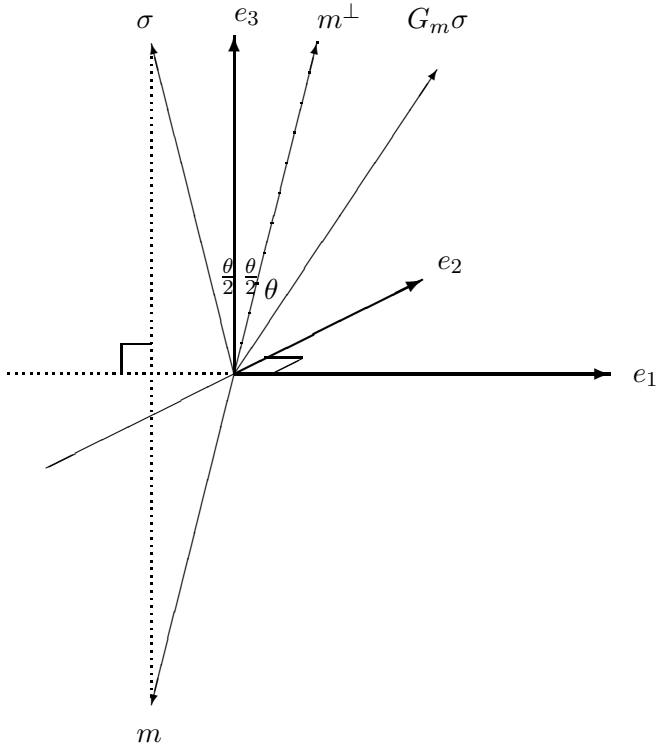
\begin{figure}[htb]
\begin{picture}(85,100)
\put(10,47){\vector(2,3){27}}
\put(10,47){\vector(1,4){11}}
\put(10,47){\vector(-1,4){11}} \put(10,47){\vector(-1,-4){11}}
\put(15,47){\line(2,1){4}}
\put(14,49.2){\line(1,0){5}}
\put(-5,47){\line(0,1){4}}
\put(-5,51){\line(1,0){4}}
\put(10,47){\line(-2,-1){25}}

\thicklines 
\put(10,47){\vector(1,0){50}} \put(10,47){\vector(0,1){45}}
\put(10,47){\vector(2,1){25}}

\put(10,94){$ \sigmaz $} \put(63,46){$ \sigmax $} \put(37,61){$ \sigmay $}
\put(-3,93){$ \sigma $} \put(33,93){$ G_m \sigma $} \put(21,93){$ m^{\perp} $}
\put(-3,-2){$ m $} 
\put(8,59){$ \frac{\theta}{2} $} 
\put(11,59){$ \frac{\theta}{2} $} 
\put(14,57){$ \theta $} 
\multiput(-1,4)(0,1){86} {\line(0,1){0.2}} 
\multiput(-20,47)(1,0){30} {\line(0,1){0.2}} 
\multiput(10,47)(1,4){12} {\line(0,1){0.2}} 
\end{picture}
\caption{Grover search in three-space based on states of maximum and minimum weight.  The two possible precession axes are now {$ \sigmax $} and {$ \sigmay $}, in order to rotate $ \sigma $ onto $ m$. }
 \label{GA3Space}
\end{figure}

\section{The Grover search operator in GA}

The action of the oracle $ G_m $ on $ | m \rangle $ is $ (I - 2 | m \rangle \langle m | ) |m\rangle = -|m\rangle $, which is to flip the `m' coordinate about the $ | m^{\perp} \rangle $ axis \cite{NielsenChuang:2002}.
Reflections are easily handled in GA, through double sided multiplication of the vector representing the axis of reflection, the action of the oracle being therefore
\be \label{eq:msigmam} 
m^{\perp} \sigma m^{\perp} =  m \sigma m .
\ee
Using the starting state defined in Eq.~(\ref{eq:mainVectors}) we find the action of the oracle on the starting state $ \sigma $ as
\be
 m \sigma m = \rme^{-\iota \sigmay \theta/2} \sigmaz \sigmaz \rme^{-\iota \sigmay \theta/4}  \rme^{-\iota \sigmay \theta/2} \sigmaz =  \rme^{-\iota \sigmay 3 \theta/2} \sigmaz = \cos \frac{3 \theta}{2} \sigmaz + \sin \frac{3 \theta}{2} \sigmax,
\ee
which is the required vector; c.f. Fig (\ref{GA3Space}).

The action of the other half of the Grover operator $ G_{\sigma} = I - 2 | \sigma \rangle \langle \sigma |  $ also produces a reflection, but this time about the $ \sigma $ vector, which therefore implies a full Grover iteration of the starting state will be
$ \sigma (m \sigma m)  \sigma = (\sigma m ) \sigma ( m \sigma) = G \sigma G^{\dagger} $, using associativity, giving the combined Grover operator as 
\be \label{eq:standardGrover}
G = - \sigma m = \rme^{\iota \sigmay \theta/2} \sigmaz \sigmaz \rme^{\iota \sigmay \theta/2} = \rme^{\iota \sigmay \theta} ,
\ee
which is a significantly more compact form for the standard Grover operator, in comparison to Eq.~(\ref{eq:StdGrover}).  We can see by inspection, that the Grover operator represents a rotation of $ 2 \theta $ about the $ \sigmay $ axis, which will clearly rotate the vector $ \sigma $ onto $ m $, after an appropriate number of operations, as shown in Fig.~\ref{GA3Space}.
Hence, after $ k $ iterations we require the Grover operator $ G $ to rotate the vector $ \sigma $, defined in Eq.~(\ref{eq:mainVectors}), onto the solution vector $ m $, and so we require
\be \label{GroverSearchEquation}
G^k \sigma {G^{\dagger}}^k = \rme^{\iota k \sigmay \theta} \rme^{\iota \sigmay \theta/2} \sigmaz \rme^{-\iota k \sigmay \theta} =  \rme^{\iota \sigmay (2 k \theta+\theta/2)} \sigmaz = m,
\ee
using $ \left ( \rme^{\iota \sigmay \theta} \right )^k = \rme^{\iota k \sigmay \theta} $, and with $ m $ defined in Eq.~(\ref{eq:mainVectors}), we therefore require
\be \label{StdGroverCondition}
\rme^{\iota \sigmay (2 k \theta+\theta/2)} \sigmaz = \rme^{\iota \sigmay (\pi-\theta/2)} \sigmaz ,
\ee
and so by equating exponents, ignoring rotations modulo $ 2 \pi $, we find the condition
\be
2 k \theta + \frac{\theta}{2} = \pi - \frac{\theta}{2},
\ee
or 
\be \label{kthetaGrover}
k =\frac{\pi}{2 \theta}-\frac{1}{2},
\ee
and using $ \theta = 2 \arcsin \sqrt{\frac{M}{N}} $ for a database with $ M $ solutions, we find
\be
k = \frac{\pi}{4 \arcsin \sqrt{\frac{M}{N}}} - \frac{1}{2} \approx \frac{\pi }{4 } \sqrt{\frac{N}{M}},
\ee
the well known result for the standard Grover search.  
However, clearly, there is no guarantee that the formula will return $ k $ as an integer, and because  it will need to be rounded to the nearest integer describing the number of Grover operations, then we will not always return exactly the solution space upon measurement.  However we can modify the search slightly, in order to guarantee that $ k $ will be an integer, and hence reliably return the solution state $ | m \rangle $.

\subsection{Exact Grover search}

The Grover operator, defined in Eq.~(\ref{eq:StdGrover}), can be modified so that it rotates the starting state  $ | \sigma \rangle $ exactly onto the solution states $ | m \rangle $, thus finding a solution with certainty \cite{genGrover,li2002general}.
In order to create the exact Grover search, the Grover operator is typically generalized to
\be
G = -(I - (1-\rme^{\rmi \phi_1}) | \sigma \rangle \langle \sigma | )(I - (1-\rme^{\rmi \phi_2}) | m \rangle \langle m | ),
\ee
so that when the oracle identifies a solution it applies a complex phase $ \rme^{\rmi \phi_2} $ to the wave function and not just the scalar $ -1 $ \cite{genGrover}.  This has the effect of slightly slowing down the search process, but it then allows the solution state $ | m \rangle $ to be reached exactly using an integral number of iterations.

A reflection can be viewed as a rotation by $ \pi $ in one higher dimension, so if we rotate by an angle $ \phi_2 $ about the $ m $ axis, which will be clockwise as viewed from above the $ \sigmaz $ axis, we obtain the oracle
\be
G_m = \rme^{\iota \frac{\phi_2}{2} (\sin (\theta/2) \sigmax + \cos (\theta/2) \sigmaz)}.
\ee
For $ \phi_2 = \pi $ we find $ G_m = \iota (\sin (\theta/2) \sigmax+\cos (\theta/2) \sigmaz) = \iota m $, so that the action of the oracle
\be
G_m \sigma G_m^{\dagger} = \iota m \sigma (-\iota m) = m \sigma m,
\ee
which gives the same result as the standard Grover oracle found previously Eq.~(\ref{eq:msigmam}),
similarly
\be
G_{\sigma} = \rme^{-\iota \frac{\phi_1}{2} (-\sin (\theta/2) \sigmax + \cos (\theta/2) \sigmaz)}
\ee
will be a rotation about the $ \sigma $ axis.
Hence for the exact search for the Grover operator we have
\bea \label{exactSearchG}
G & = & - G_{\sigma} G_m \\ \nonumber
& = & -\rme^{-\iota \frac{\phi_1}{2}(-\sin (\theta/2) \sigmax + \cos (\theta/2) \sigmaz)} \rme^{\iota \frac{\phi_2}{2}(\sin (\theta/2) \sigmax + \cos (\theta/2) \sigmaz)} , \nonumber
\eea
which when expanded gives
\bea
G & = & \cos \frac{\phi_1}{2} \cos \frac{\phi_2}{2} + \cos \theta \sin \frac{\phi_1}{2} \sin \frac{\phi_2}{2} + \sin \frac{\phi_1+\phi_2}{2} \sin \frac{\theta}{2} \iota \sigmax \\ \nonumber
&  & + \sin \frac{\phi_1}{2} \sin \frac{\phi_2}{2} \sin \theta \iota \sigmay - \cos \frac{\theta}{2} \sin \frac{\phi_1-\phi_2}{2} \iota \sigmaz  , \nonumber
\eea
for the general Grover operator, which is now interpreted as a rotation about a general precession axis.

\subsubsection{Phase matching}

We can see from Fig.~\ref{GA3Space}, which uses the alternate orthonormal basis $ | \uparrow \rangle $ and $ | \downarrow \rangle $, that $ \sigma $ and $ m $ now lie in the plane of $ \sigmax $ and $ \sigmaz $, and hence using a geometric argument the Grover precession axis must therefore lie in the plane of $ \sigmax $ and $\sigmay $ in order for the $ \sigma $ vector to be able to be rotated precisely onto the $ m $ vector.  Hence we need to eliminate the $ \sigmaz $ component in the precession axis, and hence, by inspection of Eq.~(\ref{exactSearchG}),  we require $ \phi_1 = \phi_2 $, which is the well known phase matching condition \cite{li2002general,GuiLong:1999}. Hence the exact search will be in the form
\be \label{eq:exactGphaseMatched}
G = -\rme^{\iota \beta ( \sin \alpha \sigmax + \cos \alpha \sigmay)} ,
\ee
where we find
\bea \label{eq:betaPhi}
\sin \frac{\beta}{2} & = & \sin \frac{\theta}{2} \sin \frac{\phi}{2} \\ \nonumber
\cot \alpha & = & \cos \frac{\theta}{2} \tan\frac{\phi}{2},   \nonumber
\eea
which can be re-expressed assuming a normalization factor $ Z $ as
\be
 G = \rme^{\iota \beta ( \cos \frac{\phi}{2} \sigmax + \cos \frac{\theta}{2} \sin\frac{\phi}{2} \sigmay)/Z} ,
\ee
which shows clearly the precession plane perpendicular to the vector $  \cos \frac{\phi}{2} \sigmax + \cos \frac{\theta}{2} \sin\frac{\phi}{2} \sigmay $, and if we select $ \phi = \pi $, we recover the standard Grover search operation.

To calculate $ \phi $ for the exact search we, once again have the vector equation, given by Eq.~(\ref{GroverSearchEquation}), and substituting our modified Grover operator, along with Eq.~(\ref{eq:mainVectors}), we find 
\be
\rme^{\iota k \beta( \sin \alpha \sigmax + \cos \alpha \sigmay)} \rme^{\iota \sigmay \theta/2} \sigmaz \rme^{-\iota k \beta( \sin \alpha \sigmax + \cos \alpha \sigmay)} = -\rme^{-\iota \sigmay \theta/2} \sigmaz
\ee
which can be rearranged to
\be
\rme^{\iota k \beta( \sin \alpha \sigmax + \cos \alpha \sigmay)} \rme^{\iota \sigmay \theta/2} \rme^{\iota k \beta( \sin \alpha \sigmax + \cos \alpha \sigmay)} \rme^{\iota \sigmay \theta/2}= -1
\ee
or
\be \label{finalGeneralSearchCondition}
(\rme^{\iota k \beta( \sin \alpha \sigmax + \cos \alpha \sigmay)} \rme^{\iota \sigmay \theta/2})^2 = -1.
\ee
Now, because we can always replace two consecutive precessions, with a single precession operation, we can write
\be \label{KappaEq}
\rme^{\iota k \beta( \sin \alpha \sigmax + \cos \alpha \sigmay)} \rme^{\iota \sigmay \theta/2} = \rme^{\iota \kappa \hat{\mathbf{v}}} = \cos \kappa + \iota \hat{\mathbf{v}} \sin \kappa 
\ee
for some unit vector $ \hat{\mathbf{v}} $.
Thus, from Eq.~(\ref{finalGeneralSearchCondition}), we need to solve
\bea
(\rme^{\iota \kappa \hat{\mathbf{v}} })^2 = \rme^{2 \iota \kappa \hat{\mathbf{v}} } = \cos 2 \kappa + \iota \hat{\mathbf{v}} \sin 2 \kappa  = -1  
\eea
and so clearly $\kappa = \frac{\pi}{2} $.  Thus the right hand side of Eq.~(\ref{KappaEq}), is equal to $ \iota \hat{\mathbf{v}} $, implying that the scalar part is zero.
Expanding the L.H.S. of Eq.~(\ref{KappaEq}), and setting the scalar part to zero, we find
\bea
& & \langle \left (\cos k \beta + \iota \sin k \beta  (\sin \alpha \sigmax + \cos \alpha \sigmay)\right )(\cos \frac{\theta}{2}+\iota \sin \frac{\theta}{2}  \sigmay) \rangle_0  \\
& = & \cos k \beta \cos \frac{\theta}{2}-\sin k \beta \sin \frac{\theta}{2} \cos \alpha  = 0 . \nonumber
\eea
Re-arranging this equation we find
\be
\cot k \beta = \tan \frac{\theta}{2} \cos \alpha = \frac{\sin \frac{\theta}{2}}{\sqrt{\cos^2 \frac{\theta}{2}+\cot^2 \frac{\phi}{2} }}.
\ee
Isolating $ k $, we find
\be \label{eq:kEquation}
k = \frac{\rm{arccot} \left ({\frac{\sin \frac{\theta}{2}}{\sqrt{\cos^2 \frac{\theta}{2}+\cot^2 \frac{\phi}{2} }} }\right )}{2 \arcsin(\sin \frac{\theta}{2} \sin \frac{\phi}{2} )}.
\ee
Using calculus we can find the minimum at $ \phi = \pi $, which thus returns the number of iterations for the standard Grover search given by Eq.~(\ref{kthetaGrover}), which shows that this modification fails to speed up the search \cite{zalka1999grover,Boyer:98,SSB:2005,Biham:2000,Hoyer:2000}.  However we are able now to set $ \phi $ in Eq.~(\ref{eq:kEquation}), so as to make $ k $ an integer, which will therefore be the fastest exact search possible.
Hence the minimum integer iterations will be
\be \label{eq:kmPhaseMatched}
k_m = \left\lceil \frac{\pi}{2 \theta}-\frac{1}{2} \right\rceil.
\ee
Substituting back into Eq.~(\ref{eq:kEquation}) and re-arranging we then find an expression for $ \phi $ 
\be \label{eq:exactPhaseMatchedTrans}
2 k_m \arcsin \left (\sin \frac{\theta}{2} \sin \frac{\phi}{2} \right ) = \rm{arccot} \left ({\frac{\sin \frac{\theta}{2}}{\sqrt{\cos^2 \frac{\theta}{2}+\cot^2 \frac{\phi}{2} }} } \right ),
\ee
which we can simplify to give explicitly
\be \label{eq:exactPhaseMatched}
\sin \frac{\phi}{2} =\sin \frac{\pi}{4 k_m+2} \csc \frac{\theta}{2}.
\ee
We have $ \phi $ now determined directly from the known $ \theta $ and $ k_m $ defined in Eq.~(\ref{eq:sinTheta}) and Eq.~(\ref{eq:kmPhaseMatched}) respectively, thus solving the exact search using the Grover operator defined in Eq.~(\ref{eq:exactGphaseMatched}).

An example using this formula for the exact search is given in the Appendix, which shows how the starting polarization vector now rotates exactly onto the solution states, as required.

\bigskip

\subsection{General Exact Grover search}

Most generally we can write the Grover operator as
\be
G = - U I_{\gamma} U^{-1} G_m
\ee
where
\be
I_{\gamma} = I + (\rme^{\rmi \phi_1}-1) | \gamma \rangle \langle \gamma |
\ee
where we normally choose $ \gamma = | 0 \rangle = |0 \dots 0 \rangle $, \cite{Grover:1998,Biham:99,BihKenig:2003}.  For $ U = H $ we have
\be
G = - U I_{\gamma} U^{-1} G_m = -(I +(\rme^{\rmi \phi_1}-1) H | \gamma \rangle \langle \gamma |H = -(I +(\rme^{\rmi \phi_1}-1) | \sigma \rangle \langle \sigma| =-G_{\sigma} G_m.
\ee
So with this modified operator we effectively use a modified vector to $ \sigma $, namely the vector $ \gamma = U | 0 \rangle $, giving
\bea
|\gamma \rangle & = & -\rme^{-\rmi \phi_/2} \cos \frac{\theta_0}{4} |\uparrow \rangle + \rme^{\rmi \phi/2} \sin \frac{\theta_0}{4} |\downarrow \rangle ,  \nonumber
\eea
equivalent to a starting polarization vector
\be
\gamma  =  -\sin \frac{\theta_0}{2} \cos \phi_0 \sigmax - \sin \frac{\theta_0}{2} \sin \phi_0 \sigmay + \cos \frac{\theta_0}{2}  \sigmaz .
\ee
Comparing this with the polarization vector for the standard Grover search  $ \sigma = -\sin \frac{\theta}{2} \sigmax +\cos \frac{\theta}{2} \sigmaz $, as shown on Fig~\ref{GA3Space}, we see that we have changed the projection in the $ \sigmaz $ direction by changing $ \theta $ to $ \theta_0 $, and hence rotated the vector in the $ e_{12} $ plane given by the vector $ \phi_0 $.  If $ \phi_0 = 0 $, then we recover the standard exact Grover search. As this is a unit vector, we simply adapt $ G_{\sigma} $ to rotate about this new vector, that is we have
\be
G_{\gamma} = \rme^{-\iota \gamma \phi_1/2 } = \rme^{-\iota \frac{\phi_1}{2} ( -\sin \frac{\theta_0}{2} \cos \phi_0 \sigmax - \sin \frac{\theta_0}{2}  \sin \phi_0 \sigmay+\cos \frac{\theta_0 }{2} \sigmaz)}
\ee
and hence for the general exact search we have
\be \label{generalGroverOp}
G = - G_{\gamma} G_m = \rme^{-\iota \frac{\phi_1}{2} ( -\sin \frac{\theta_0}{2} \cos \phi_0 \sigmax - \sin \frac{\theta_0}{2}  \sin \phi_0 \sigmay+\cos \frac{\theta_0 }{2} \sigmaz)} \rme^{\iota \phi_2/2(\sin (\theta/2) \sigmax + \cos (\theta/2) \sigmaz)}.
\ee
However, as a more elegant alternative, we can simply adjust our basis states, given by Eq.~(\ref{maxminWeight}), and then the exact solution, given by Eq.~(\ref{eq:exactGphaseMatched}), immediately applies.

\section{Summary}

The Grover search algorithm is a central algorithm in the field of quantum computing, and hence it is important to represent it in the most efficient formalism possible. The two main strengths of geometric algebra are its method of handling rotations and its integral geometric representation, and hence its perfect suitability in describing the Grover search.  
We find Clifford's geometric algebra, provides a simplified representation for the Grover operator Eq.~(\ref{eq:standardGrover}) and a clear geometric picture of the search process.  Using the states of maximum and minimum weight, we find that we can interpret the  search process as the precession of a spin-$\frac{1}{2}$ particle, thus providing a simple visual picture, as shown in Fig~\ref{GA3Space}. This is not possible with the standard formalism as it requires two complex axes, forming a four-dimensional space, and hence difficult to visualize. 
We also find that the exact Grover search Eq.~(\ref{eq:exactGphaseMatched}) has an efficient algebraic solution, as shown in Eq.~(\ref{eq:exactPhaseMatched}). 
Improved intuition obtained via the use of Clifford's geometric algebra, may possibly enhance the search for new quantum algorithms.

\appendix

\section{Example of an exact search over 16 elements}

After $ k $ iterations we have the polarization vector
\bea
P & = & G^k \sigma {G^{\dagger}}^k \\ \nonumber
 & = & \rme^{\iota k \beta (\sin \alpha \sigmax+\cos \alpha \sigmay} \rme^{\iota \sigmay \theta/2} \sigmaz \rme^{-\iota k \beta (\sin \alpha \sigmax+\cos \alpha \sigmay} \\ \nonumber
& = &  - \left (\sin^2 \alpha \sin \frac{\theta}{2}+ \sin \frac{\theta}{2} \cos^2 \alpha \cos 2 \beta k + \cos \alpha \cos \frac{\theta}{2} \sin 2 \beta k \right ) \sigmax \\ \nonumber
& + & \left (-\frac{1}{2} \sin \frac{\theta}{2} \sin 2 \alpha + \frac{1}{2} \sin 2 \alpha \sin \frac{\theta}{2} \cos 2 \beta k + \cos \frac{\theta}{2} \sin \alpha \sin 2 \beta k \right ) \sigmay \\ \nonumber
& + & \left ( \cos \frac{\theta}{2} \cos 2 \beta k - \cos \alpha \sin \frac{\theta}{2} \sin 2 \beta k \right ) \sigmaz . \nonumber
\eea
For 16 elements we find from Eq.~(\ref{eq:kmPhaseMatched}) $ k_m = 3 $, and we then find $ \phi $ from Eq.~(\ref{eq:exactPhaseMatched}) for an exact search of $ \phi = 2.19506 $.
This gives the polarization vector after $ k $ iterations 
\bea
P & = &  -(0.0546434 + 0.195357 \cos2 \beta k + 0.855913 \sin 2 \beta k ) \sigmax \\ \nonumber
& + & (-0.10332 + 0.10332 \cos 2 \beta k + 0.452673 \sin 2 \beta k) \sigmay \\ \nonumber
& + & (0.968246 \cos 2 \beta k - 0.220996 \sin 2 \beta k) \sigmaz . \nonumber
\eea
Using $ \alpha $ and $ \beta $ defined in Eq.~(\ref{eq:betaPhi}), beginning from a starting vector $ \sigma = ( -0.25,0,0.9682 ) $, with a required solution vector $ m = ( -0.25,0,-0.9682 ) $, we generate a polarization vector~$ P $ as
\bea
\sigma & = &  ( -0.25,0,0.9682 ) \\ \nonumber
G \sigma & = &  ( -0.8456,0.315,0.4309 ) \\ \nonumber
G^2 \sigma & = &  ( -0.8456,0.315,-0.4309 ) \\ \nonumber
G^3 \sigma & = &  ( -0.25,0,-0.9682 ) \nonumber
\eea
thus producing the exact solution $ m $ after $ k_m = 3 $ iterations as required.

%\bibliography{oneGrover}% Produces the bibliography via BibTeX.
\bibliography{quantum}% Produces the bibliography via BibTeX.

\end{document}